# Acceleration of DNA Replication of Klenow Fragment by Small Resisting Force

Yu-Ru Liu(刘玉如)[1], Peng-Ye Wang(王鹏业)[1,2,3], Wei Li(李伟)[1,3*], and Ping Xie(谢平)[1*]

[1]*Laboratory of Soft Matter Physics and Beijing National Laboratory for Condensed Matter Physics, Institute of Physics, Chinese Academy of Sciences, Beijing 100190, China*

[2]*School of Physical Sciences, University of Chinese Academy of Sciences, Beijing 100049, China*

[3]*Songshan Lake Materials Laboratory, Dongguan 523808, China*

*Corresponding authors. Email: weili007@iphy.ac.cn; pxie@iphy.ac.cn

Supported by the Natural Science Foundation of China (11674381, 21991133, 11774407, 11874415, 11874414, 31770812), Key Research Program on Frontier Science (QYZDB-SSWSLH045), National Key Research and Development Program (2016YFA0301500), CAS Strategic Priority Research Program (XDB37010100), and National Laboratory of Biomacromolecules (2020kf02).

**Abstract**

DNA polymerases are an essential class of enzymes or molecular motors that catalyze processive DNA syntheses during DNA replications. A critical issue for DNA polymerases is their molecular mechanism of processive DNA replication. We have previously proposed a model for chemomechanical coupling of DNA polymerases, based on which the predicted results have been provided about the dependence of DNA replication velocity upon the external force on Klenow fragment of DNA polymerase I. Here, we performed single molecule measurements of the replication velocity of Klenow fragment under the external force by using magnetic tweezers. The single molecule data verified quantitatively the previous theoretical predictions, which is critical to the chemomechanical coupling mechanism of DNA polymerases. A prominent characteristic for the Klenow fragment is that the replication velocity is independent of the assisting force whereas the velocity increases largely with the increase of the resisting force, attains the maximum velocity at about 3.8 pN and then decreases with the further increase of the resisting force.

DNA polymerases are central enzymes in the DNA replication process. They function as molecular motors that can catalyze processive DNA syntheses by translocating along DNA template. A well-studied example of them is Klenow fragment of *Escherichia coli* DNA polymerase I, which is an active truncated form composed of a polymerase domain and a 3'-5' exonuclease domain.[1–3] An important issue for the DNA polymerase is how it makes processive DNA replication. To address the issue, besides extensive structural, biochemical and single molecule studies, theoretical modeling and molecular dynamics simulations have also attracted much attention. For example, structural studies showed that the polymerase domain of the enzyme is composed of three subdomains: the fingers, palm and thumb.[4–8] Comparison of the structures of DNAP-DNA binary complexes with the corresponding DNAP-DNA-dNTP ternary complexes showed a large rotation of the fingers relative to the palm and thumb.[9–13] The rate constants of the fingers rotations were measured biochemically.[14,15] The translocation of Klenow fragment along the template during DNA replication was monitored using single molecule fluorescence resonance energy transfer (smFRET) with single base-pair resolution.[16] Moreover, using smFRET the dynamics of strand displacement DNA replication by Klenow fragment was also studied.[17] In particular, using single molecule optical or magnetic trapping techniques, the replication velocity was studied elaborately as the DNA polymerase catalyzes the replication of a mechanically stretched DNA template.[18,19] Correspondingly, theoretical modeling studies[20,21] and molecular dynamics simulations[22,23] have been presented to explain these single molecule data. These studies advanced greatly our understanding of the chemomechanical coupling mechanism of DNA polymerases. On the other hand, knowing how the external force acting directly on the enzyme on the DNA replication is also implicated to the working mechanism of the enzyme. As a result, using theoretical modeling Xie has studied the dynamics of Klenow fragment under the external force on it.[24] It was predicted interestingly that while the force assisting the downstream translocation of the enzyme has little effect on the DNA replication rate, a small force resisting the downstream translocation can enhance largely the replication rate.[24] However, no experimental data are available on the effect of the external force acting directly on the enzyme on the dynamics of DNA replication. Testing the theoretical predictions is critical to the chemomechanical coupling mechanism of DNA polymerase. For this purpose, in this work we used single molecule magnetic trapping to measure the replication dynamics of Klenow fragment under the external force acting directly on the enzyme.

In Fig. 1(a) we interpret schematically the model for the processive DNA replication by the high-fidelity DNA polymerase.[24] In the model, it is proposed that after a nucleotide (dNTP) binding to the polymerase active site the fingers rotate inwards and towards the active site, with the fingers transiting from open to closed conformation [Fig. 1(a)]. After incorporation of dNTP the fingers rotate outwards away from the active site, with the fingers transiting to the open conformation [Fig. 1(a)]. These fingers rotations are consistent with the structural data.[9–13] After incorporation of dNTP paired with a DNA base on the template, the downstream translocation of DNA polymerase arises from the change in the binding energy of the enzyme to the DNA substrate,[24–27] as stated briefly as follows. The interaction between the enzyme and DNA substrate can be characterized by two DNA-binding sites in the enzyme. One binding site (called *site I*) contains residues in the fingers that can interact mainly with the 3' – 5' single-stranded DNA (ssDNA) [Fig. 1(a)], which is consistent with the biochemical evidence.[28,29] The other binding site (called *site II*) contains residues in palm and thumb that can interact mainly with double-stranded DNA (dsDNA) [Fig. 1(a)]. For convenience, we represent the position of the enzyme by that of its active site along the DNA substrate. When the enzyme is positioned at the first unpaired base on the 3' – 5' ssDNA which is counted from the dsDNA region, the enzyme is denoted as at the $n$th position. When the enzyme is positioned at the second unpaired base on the 3' – 5' ssDNA, the enzyme is denoted as at the ($n$+1)th position. When the enzyme is positioned at the first base pair that is counted from the ssDNA region, the enzyme is denoted as at the ($n$–1)th position. If the enzyme is at the $n$th position, all residues in *site I* can interact with ssDNA and all residues in *site II* can interact with dsDNA, and thus the enzyme has the maximum binding energy to the DNA substrate [inset of Figs. 1(a) and 1(b)]. By contrast, if the enzyme is at the ($n$–1)th position, although all residues in *site II* can interact with dsDNA only some residues in *site I* can interact with ssDNA, and thus the enzyme has a smaller binding energy to the DNA substrate than at the $n$th position [inset of Figs. 1(a) and 1(b)]. Similarly, if the enzyme is at the ($n$+1)th position, although all residues in *site I* can interact with ssDNA only some residues in *site II* can interact with dsDNA, and thus the enzyme has a smaller binding energy to the DNA substrate than at the *n*th position [inset of Figs. 1(a) and 1(b)]. Hence, the enzyme is most of time in the *n*th position, with the active site being at the first unpaired base on the 3' – 5' ssDNA template.

To study the effect of the force on the replication dynamics, consider an external force acting directly on the enzyme, with the force being defined as positive if it is toward the 5' end of the 3' – 5' ssDNA template, i.e., along the translocation direction of the enzyme, and as negative if it is in the opposite direction. The negative force can be realized by applying the force on the enzyme and with the 5'-end of the ssDNA template being fixed [see Fig. 2(a)], while the positive force can be realized by applying the force on the enzyme and with the upstream end of dsDNA being fixed [see Fig. 2(c)]. First, consider the negative force. As the fingers are bound strongly to the 3' – 5 ssDNA, the inward rotation of the fingers would drive the magnetic bead attached to the enzyme to move a distance *d* against the external force and the outward rotation would drive the bead to make the same distance *d* in the reverse direction. In other words, the negative force resists the inward rotation of the fingers while facilitates the outward rotation. Based on the model, the force dependence of the replication rate can be approximately written as [24]

$$k(F) = \frac{k_c(F)[\text{dNTP}]}{[\text{dNTP}] + K_m(F)\left[\exp\left(\frac{\Delta E}{k_BT}\right)\exp\left(-\frac{Fp}{k_BT}\right)+1\right] \Big/ \left[\exp\left(\frac{\Delta E}{k_BT}\right)+1\right]}, \qquad (1)$$

where $F$ is the external force, $k(F)$ is the replication rate, which is dependent on $F$, [dNTP] is dNTP concentration, $p = 0.34$ nm is the distance between two base pairs, $\Delta E$ is the difference of the binding energy of the enzyme to the DNA substrate at the (*n*−1)th position and that at the *n*th position, and $k_BT$ is the Boltzmann constant times the absolute temperature. In Eq. (1), $k_c(F)$ approximately has the form [24]

$$\frac{1}{k_c(F)} = \frac{1}{k_1} + \frac{1}{k_2(F)} + \frac{1}{k_3(F)}, \qquad (2)$$

where $k_1$ is the rate of the ternary complex of the enzyme, DNA substrate and dNTP transition from the inactivated to activated state after dNTP binding, which is independent of $F$, $k_2(F)$ is the rate of the inward rotation of the fingers after the ternary complex transition to the activated state, which is dependent on $F$, and $k_3(F)$ is the rate of the outward rotation of the fingers after the incorporation of dNTP, which is also dependent on $F$. The two rates $k_2(F)$ and $k_3(F)$ have following Arrhenius-Eyring forms [24]

$$k_2(F) = k_{20} \exp\left(\frac{Fd}{k_B T}\right), \tag{3a}$$

$$k_3(F) = k_{30} \exp\left(-\frac{Fd}{k_B T}\right), \tag{3b}$$

where $k_{20}$ and $k_{30}$ are rates of $k_2(F)$ and $k_3(F)$ under no force. $K_m(F)$ in Eq. (1) has the form $K_m(F) = k_c(F)/k_b$, where $k_b$ is the second-order rate of dNTP binding, which is independent of $F$. Second, consider the positive force. As it is noted, the force has no effect on the rotation of the figures. Thus, the replication rate can still be calculated by Eq. (1), but with $k_c(F)$ being replaced with $k_c(0) = (1/k_1 + 1/k_2(0) + 1/k_3(0))^{-1}$ and $K_m(F)$ being replaced with $K_m = k_c(0)/k_b$ independent of $F$.[24]

From Eqs. (1) – (3), it is noted that to calculate the force dependence of replication rate it is required to know values of parameters $k_b$, $k_1$, $k_{20}$, $k_{30}$, $\Delta E$ and $d$. As mentioned before,[24] the available biochemical data for Klenow fragment gave $k_b = 13$ μM$^{-1}$s$^{-1}$,[14] $k_1 = 200$ s$^{-1}$,[14] $k_{20} = 200$ s$^{-1}$ [14] and $k_{30} = 15$ s$^{-1}$.[15] From the available biochemical data[28] it was deduced $\Delta E < -2.6 k_B T$.[24] As determined before,[24] $d = 1.6$ nm, which is consistent with the estimation from the available structural data.[11] With above parameter values, the predicted results of the replication velocity versus $F$ for different values of $\Delta E$ are shown in Fig. 1(c).[24] As it is noted, the results are nearly independent of $\Delta E$ provided that $\Delta E < -2.6 k_B T$.[24]

To test the above prediction of the replication velocity versus $F$, we employed single molecule magnetic tweezers. The experimental configuration is schematically shown in Fig. 2(a,c). The Klenow fragment was biotin tagged by adding the amino acid sequence MAGGLNDIFEAQKIEWHE at the N terminus (the underlined Lysine is biotinylated). The tag is located in the palm domain, which is far away from the active site and thus does not affect the polymerase activity. For the case of resisting force, template ssDNA was PCR amplified with 5'-Dig-tagged primer. For the case of assisting force, multiple-Dig-tagged short dsDNA fragment was PCR amplified with Dig-dUTP and then ligated into 3'-terminus of template DNA. The experimental procedure is described as follows. The isolated ssDNA templates were re-annealed with primers. The diluted DNA substrates were then allowed to assemble with Bio-tagged Klenow at equal molar ratio (10 nM for both) at 25℃ for 10–30 min in the buffer (50 mM NaCl, 10 mM Tris-HCl, 10 mM $MgCl_2$, 1 mM dithiothreitol). Subsequently, the complex was adjusted to 2 pM, injected into flow cell (~50 μl in volume) and incubated for 30 min at 25℃. 100 μl streptavidin-coated magnetic beads (1 μm in diameter, ~$10^6$/ml) were injected. After a brief incubation (5–10 min), the flow cell was washed with the buffer to remove the free beads, and the buffer with the addition of 500 μM dNTPs (with equal concentration of dATP, dGTP, dCTP and dTTP) were injected into the flow cell for initiating the replication. The measurements were undertaken under constant applied force mode. The force ranged from 0.4 pN to 11.3 pN for both assisting and opposing directions.

Some trajectories of the change in the distance of the magnetic bead are plotted in Fig. 2(b,d). From a measured trajectory a replication velocity was computed by dividing the total distance of advance by the total elapsed time during which the distance is increasing or decreasing. For a given force, the velocity shows a large variation, which can be seen from Fig. 2(e), where distributions of velocities ($N = 60 - 300$) for some external forces are shown. The large variation could reflect differences in polymerase activity among individual molecules, similar to those observed in prior single molecule experiments for DNA polymerases[18,19] and for other enzymes[30,31]. The arithmetic average velocities under different forces are shown in Fig. 3 (dots). Remarkably, it is seen that with parameter values determined in the literature,[14,15] as given in Fig. 1c, the theoretical results (dashed line in Fig. 3) are consistent with our single molecule data. Due to different enzyme preparations and different buffer conditions, the parameter values in our assay could be slightly different from those measured by other researcher groups.[14,15] If we adjust $k_1 = 180$ $s^{-1}$, $k_{20} = 180$ $s^{-1}$ and $k_{30} = 17$ $s^{-1}$ while take the same values for other parameters as given in Fig. 1(c), the theoretical results (solid line in Fig. 3) become better agreement with our singe molecule data. A prominent feature of both the experimental and theoretical results is that a small resisting force can enhance largely the DNA replication velocity, which is contrary to the common sense that the resisting force would slow the replication. The quantitative agreement between the experimental and predicted theoretical results strongly supports the validity of the model.

In summary, using single molecule magnetic tweezers we measured DNA

replication velocity by Klenow fragment under both assisting and resisting forces on the enzyme. The measured force-velocity curves verify quantitatively the previous theoretical predictions that the velocity is nearly independent of the assisting force whereas the velocity increases largely with the increase in the magnitude of the resisting force, attains the maximum velocity at about 3.8-pN resisting force and then decreases with the further increase in the magnitude of the resisting force. The present single molecule studies thus give a strong support to the previous model, advancing significantly our understanding of the molecular mechanism of the chemomechanical coupling of DNA polymerases. Moreover, it is interesting to note that a small resisting force can facilitate the DNA replication, which may be important to the biological function of the DNA polymerases.

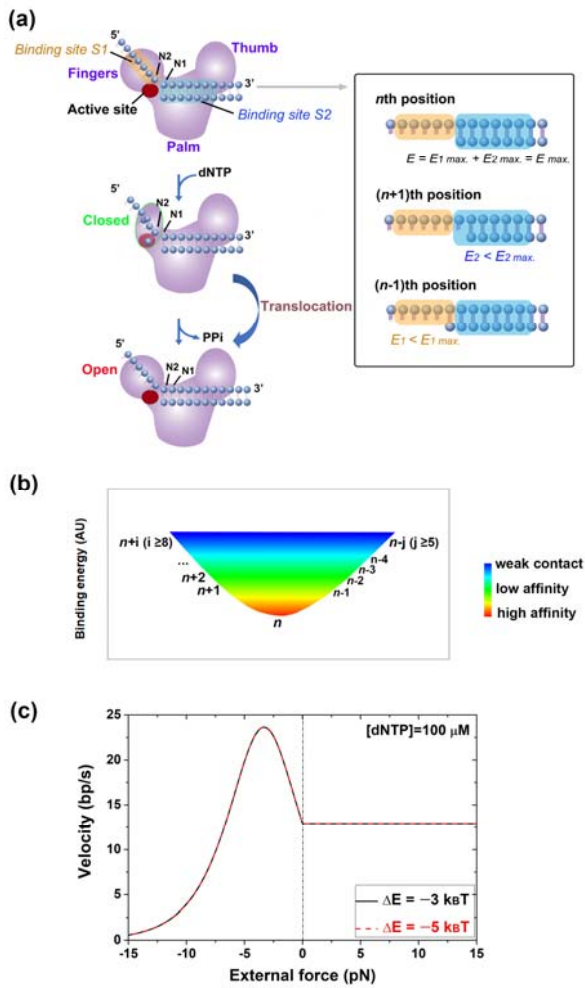

Fig. 1. The proposed model for DNA polymerase. (a) The pathway of DNA replication. Inset illustrates the definition of the position of the enzyme relative to the DNA substrate. (b) Schematic illustration of the relative binding energy of the enzyme to the DNA substrate. (c) Predicted results of replication velocity versus external force for Klenow fragment at nearly saturating concentration of dNTP (100 μM). The data are reproduced from Ref. [24], which were calculated with parameter values determined in the literature (see text). The assisting force is defined to be positive and the resisting force is defined to be negative. $\Delta E$ is the difference of the binding energy of the enzyme to the DNA substrate at the $(n-1)$th position and that at the $n$th position.

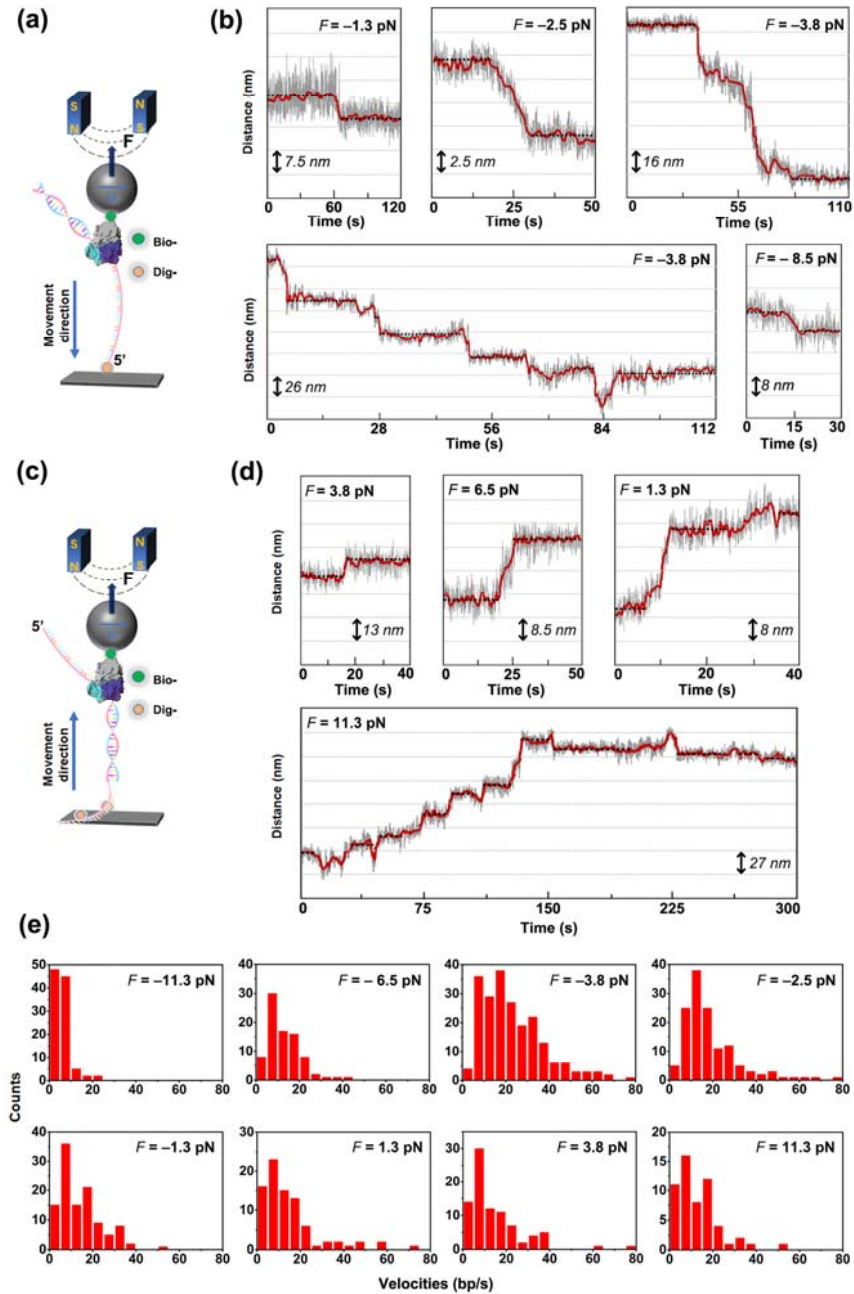

Fig. 2. Direct measurements of DNA replication velocity of Klenow fragment under the external force. (a) Experimental configuration for the case of resisting force. Klenow fragment is attached to the magnetic bead through a Biotin-tag at N terminus. Template ssDNA (3kb) was immobilized by 5'-Dig tags onto the surface of the coverslip. (b) Some trajectories of the change in the distance of the magnetic bead under different values of the resisting force, which is defined to be negative. (c) Experimental configuration for the case of assisting force. Template ssDNA was multiple-Dig-tagged at 3'-terminus and attached to the surface of the coverslip. (d) Some trajectories of the change in the distance of the magnetic bead under different values of the assisting force, which is defined to be positive. (e) Distributions of velocities under some external forces.

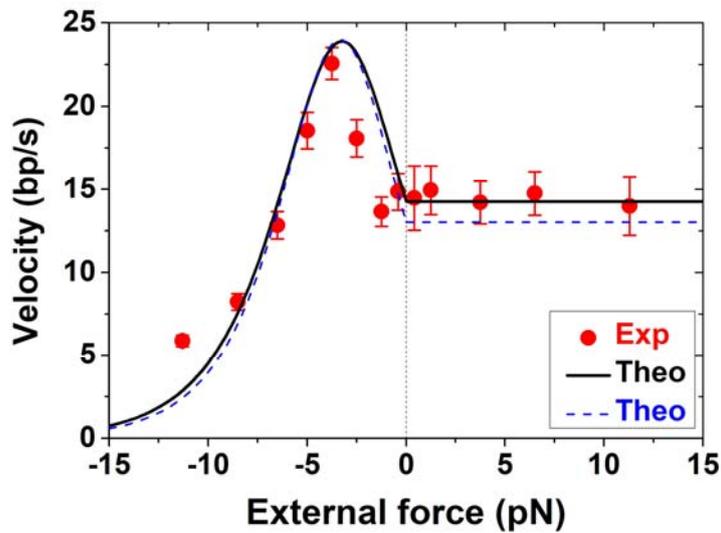

Fig. 3. Replication velocity versus external force. The assisting force is defined to be positive and the resisting force is defined to be negative. Lines are theoretical results, with dashed line being calculated with parameter values determined in the literature, as used in Fig. 1(c), and solid line being calculated with $k_1 = 180$ $s^{-1}$, $k_{20} = 180$ $s^{-1}$, $k_{30} = 17$ $s^{-1}$ and values of other parameters as given in Fig. 1(c). Dots are single molecule data. Under the assisting force, the replication velocity is about 14 bp/s, which is close to the prior biochemical data (15bp/s) [15]. The error bar is the "error of the mean", i.e. (standard deviation) $\times$ $N^{-1/2}$, where $N$ is the number of data points taken to obtain the average value (with $N = 60 - 300$ in our measurements).